\documentclass{article}

\usepackage[preprint]{neurips_2026}

\usepackage[utf8]{inputenc} %
\usepackage[T1]{fontenc}    %
\usepackage{hyperref}       %
\usepackage{url}            %
\usepackage{booktabs}       %
\usepackage{amsfonts}       %
\usepackage{nicefrac}       %
\usepackage{microtype}      %
\usepackage{xcolor}         %
\usepackage{amsmath}
\usepackage{bm}
\usepackage{comment}
\usepackage{graphicx}
\usepackage{enumitem}

\usepackage{listings}

\usepackage{listings}
\usepackage{xcolor}

\makeatletter
\renewcommand{\paragraph}{%
  \@startsection{paragraph}{4}%
  {\z@}{1ex \@plus .5ex \@minus .2ex}{-1em}%
  {\normalfont\normalsize\bfseries}%
}
\makeatother

\def\vtheta{{\bm{\theta}}}

\def\vc{{\bm{c}}}
\def\vd{{\bm{d}}}
\def\ve{{\bm{e}}}
\def\vf{{\bm{f}}}

\def\vh{{\bm{h}}}

\def\vj{{\bm{j}}}

\def\vn{{\bm{n}}}

\def\vp{{\bm{p}}}

\def\vt{{\bm{t}}}

\def\vy{{\bm{y}}}
\def\vz{{\bm{z}}}

\def\evq{{q}}

\def\mE{{\bm{E}}}

\def\mH{{\bm{H}}}
\def\mI{{\bm{I}}}

\def\mO{{\bm{O}}}
\def\mP{{\bm{P}}}

\def\mW{{\bm{W}}}

\def\mZ{{\bm{Z}}}

\newcommand{\method}{\textcolor{blue}{Rigel3D}} 
\title{\method: 
\textcolor{blue}{Rig}-awar\textcolor{blue}{e} 
\textcolor{blue}{L}atents 
for\\Animation-Ready \textcolor{blue}{3D} Asset Generation}

\usepackage{authblk}

\author{
    \textbf{Nikitas Chatzis\textsuperscript{1},}
    \textbf{Marios Loizou\textsuperscript{1,2},}
    \textbf{Evangelos Kalogerakis\textsuperscript{1,2,3}}\\
    \textsuperscript{1}Technical University of Crete \quad \textsuperscript{2}CYENS Center of Excellence \\ \textsuperscript{3}University of Massachusetts Amherst
}

\begin{document}

\maketitle

\let\oldthefootnote\thefootnote
\let\thefootnote\relax
\footnotetext{\noindent$^{\dag}$ In accordance with the ERC Open Access mandate, the authors have made 
the Author Accepted Manuscript (AAM) publicly available under the Creative Commons Attribution (CC-BY 
4.0) license.}
\let\thefootnote\oldthefootnote

\begin{abstract}
Recent 3D generative models can synthesize high-quality assets, but their outputs are typically static: they lack the skeletal rigs, joint hierarchies, and skinning weights required for animation. This limits their use in games, film, simulation, virtual agents, and embodied AI, where assets must not only look plausible but also move plausibly. We introduce \method{}, a generative method for animation-ready 3D assets represented as rigged meshes. Unlike post-hoc auto-rigging methods that attach rigs to completed shapes, our method jointly models geometry and rig structure through coupled surface and skeleton structured latent representations. A rig-aware autoencoder decodes these representations into mesh geometry, skeleton topology, joint coordinates, and skinning weights, while a two-stage latent generative model synthesizes both surface and skeleton representations for image-conditioned generation. To support downstream animation workflows, we further introduce an open-vocabulary joint labeling module that embeds generated joints into a shared vision-language space, enabling correspondence to arbitrary retargeting templates. Experiments on large-scale rigged asset datasets demonstrate that our method generates diverse, high-quality animation-ready assets and outperforms existing rigging baselines across multiple metrics.
\end{abstract}

\section{Introduction}
\label{sec:intro}

Recent advances in 3D generative modeling have made it possible to synthesize increasingly detailed 3D assets from images, text, and learned latent distributions~\cite{zhang2023vecset,zhang2024clay,xiang2024trellis,
xiang2025trellis2}. Yet most generated assets remain static: they lack the skeletal rigs, joint hierarchies, and skinning weights required by standard animation pipelines. Consequently, a generated mesh may look plausible but still require substantial post-processing before it can be posed, animated, or retargeted to existing motion data.
Automatic rigging has long aimed to bridge this gap by predicting skeletons and skinning weights for input meshes. 
Classical and learning-based methods such as Pinocchio~\cite{baran2007pinnochio},  RigNet~\cite{xu2020rignet}, TARig~\cite{ma2023tarig}, and recent template-free approaches~\cite{deng2025anymate,Song2025magicarticulate,song2025puppeteer} have made significant progress in turning static shapes into animation-ready assets. 
However, most of these methods treat rigging as a post-processing step applied after shape generation. 
This separation is limiting for generative outputs, where geometry, topology, pose, and part structure may differ substantially from the training distribution of a downstream rigger. 
Instead of first generating a static mesh and then attaching a rig, we seek to generate geometry and rigging together.

\begin{figure}
    \centering
    \includegraphics[width=\textwidth]{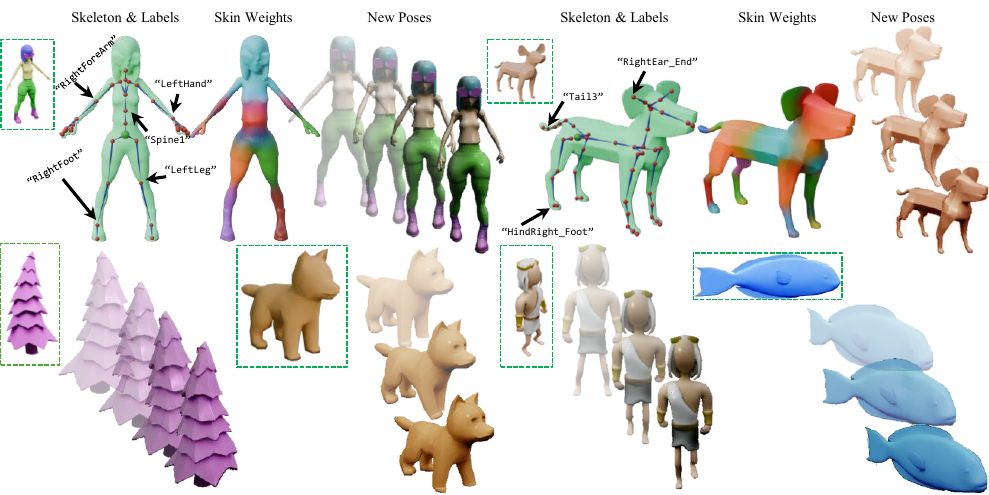}
    \vspace{-18pt} 
    \caption{Given input images (in green boxes), \method{} generates diverse rigged 3D assets with meshes, textures, skeletons, labeled joints, and skinning weights (top row), making them directly usable in standard animation pipelines. Bottom row shows animated poses only.
     } 
    \label{fig:teaser}
    \vspace{-15pt}    
\end{figure}

This joint generation problem is challenging because the relevant structures are heterogeneous but tightly coupled. 
Surface geometry is naturally represented by local features near the visible surface, while skeletons are sparse hierarchical structures that predominantly lie  inside the object. 
A model must preserve consistency between these two domains, generate valid variable-size kinematic trees, and predict continuous skinning weights that bind surface vertices to bones. 
Another challenge is that practical animation workflows often require semantic joint labels for motion retargeting, but template-free generated skeletons typically provide only coordinates and connectivity. 
Thus, animation-ready asset generation requires not only high-quality shape synthesis, but also coherent skeleton generation, skinning prediction, and semantic compatibility with existing animation tools.

We introduce \method{}, a generative method for animation-ready 3D asset synthesis. 
Our method builds on TRELLIS' Structured LATent representation (SLATs) ~\cite{xiang2024trellis}, that represent  3D assets using sparse voxel locations and local latent codes attached to those voxels. 
We extend this surface-centric representation to rig-aware generation by learning two coupled structured latent representations: a surface SLat capturing geometry and appearance, and a skeleton SLat capturing articulation structure. 
A rig-aware autoencoder maps rigged meshes into these latent representations and decodes them into mesh geometry, skeleton topology, joint coordinates, and skinning weights. 
The decoder combines a skeleton-conditioned mesh decoder, an autoregressive skeleton decoder, and an attention-based skinning decoder, allowing shape and rig structure to inform one another.
We then train a generative model over the rig-aware SLats. The generated latent representations are decoded into a mesh, optional appearance, an animation skeleton, and skinning weights, yielding complete rigged 3D assets rather than static ones. 
To support downstream motion retargeting, we also introduce an open-vocabulary joint labeling module that embeds generated joints into a shared vision-language space. Unlike closed-set classifiers tied to a fixed template, this module allows a generated skeleton to be matched to arbitrary candidate label sets. In summary, our contributions are:

\begin{itemize}[leftmargin=*, itemsep=0.2em, topsep=0.2em]
    \item We introduce \method{}, a generative end-to-end framework for animation-ready 3D assets that produces meshes, optional appearance, skeletons, and skinning weights.
    \item We propose a rig-aware autoencoder with coupled surface and skeleton SLats, a skeleton-conditioned mesh decoder, an autoregressive skeleton decoder, and an attention-based skinning decoder, along with two-stage SLat generation for both surface and skeleton latent representations, enabling joint generation of geometry and rig structure.
    \item We introduce an open-vocabulary joint labeling module that supports motion retargeting by matching generated joints to text labels in a shared vision-language space.
    \item We demonstrate state-of-the-art performance over prior auto-rigging baselines across several metrics on two datasets: Anymate~\cite{deng2025anymate} and ModelsResource~\cite{xu2020rignet}.
\end{itemize}

\section{Related Work}
\label{sec:related_work}

\paragraph{3D and 4D generation.}
Recent 3D generative models synthesize high-quality static assets from images, text, or latent distributions using neural fields, Gaussian splats, vector-set representations, and sparse structured latents~\cite{mildenhall2020nerf,kerbl3Dgaussians,zhang2023vecset,zhang2024clay,xiang2024trellis}.
TRELLIS~\cite{xiang2024trellis} introduced SLats for scalable two-stage 3D generation, with follow-up work improving compactness and fidelity~\cite{xiang2025trellis2}. 
Other methods generate dynamic content or articulated objects~\cite{ren2023dreamgaussian4d,wu2025animateanymesh,chen2025ArtiLatent,li2025particulate}, but typically represent motion, rigid-only articulation, or deformation sequences rather than complete skeletal rigs with skinning weights. 
Our method instead generates rigged meshes that can be controlled through standard skeletal animation pipelines.

\paragraph{Automatic rigging.}
Automatic rigging aims to convert a static 3D shape into an animation-ready asset by predicting an internal skeleton and skinning weights. 
Classical methods such as Pinocchio~\cite{baran2007pinnochio} embed template skeletons into input meshes, while early learning-based rigging methods such as volumetric skeleton prediction~\cite{xu2019predicting} and RigNet~\cite{xu2020rignet} predict skeleton structure from 3D geometry. 
Several works study humanoid, character-specific, or template-aware rigging: Neural Blend Shapes~\cite{li2021neuralblendshapes} assumes a prescribed skeleton structure and learns pose-dependent corrective deformations, NeuroSkinning~\cite{liu2019neuroskinning} predicts skinning weights for production characters with known skeletons, TARig~\cite{ma2023tarig} performs template-aware humanoid rigging, and MoRig~\cite{xu2022morig} uses motion cues from point-cloud sequences to infer rigs. 
Recent humanoid-oriented systems such as HumanRig~\cite{chu2024humanrig}, DRiVE~\cite{sun2024drive}, and Make-It-Animatable~\cite{guo2025animatable} produce animation-ready characters, but are primarily designed for humanoid inputs or post-hoc rigging of given shapes. 
CANOR~\cite{he2025canor} takes a different approach by predicting editable control blobs instead of an explicit skeleton and skinning representation.

More recent template-free auto-rigging methods improve generality by modeling skeletons autoregressively or decomposing rigging into joint, connectivity, and skinning stages. 
Anymate~\cite{deng2025anymate} introduces a large rigged-object dataset and strong modular baselines for joint prediction, connectivity, and skinning. 
MagicArticulate~\cite{Song2025magicarticulate} and Puppeteer~\cite{song2025puppeteer} use transformer-based skeleton generation conditioned on shape features, while UniRig~\cite{zhang2025unirig}, RigAnything~\cite{liu2025riganything}, ARMO~\cite{sun2025armo}, Auto-Connect~\cite{guo2025autoconnectconnectivitypreservingrigformerdirect} and \cite{sun2026animatorcentricskeletongenerationobjects} explore autoregressive skeleton tokenization, connectivity modeling, and skinning prediction for diverse assets. 
These methods generally assume a completed input shape and infer a rig as a post-processing step. 
In contrast, our method jointly models the interdependent shape geometry and rig structure in an end-to-end generative framework.

\paragraph{Generative animation-ready assets.}
Closest to our work are methods that synthesize assets together with animation structure. 
SKDream~\cite{xu2025skdream} conditions multiview and 3D generation on arbitrary skeletons, but assumes a skeleton input rather than generating a complete rig from scratch. 
AnimaX~\cite{huang2025animax}, AnimaMimic~\cite{xie2025animamimic}, Make-It-Poseable~\cite{guo2025makeitposeable}, and AnimateAnyMesh~\cite{wu2025animateanymesh} animate existing meshes or generate motion-conditioned deformations, but do not primarily focus on jointly generating a new mesh, skeleton, and skinning weights as a complete rigged asset. 
AnyTop~\cite{gat2025anytop} generates motions for arbitrary skeleton topologies through textual joint descriptions. 
AniGen~\cite{huang2026anigen} is concurrent work that directly generates animatable 3D assets by representing shape, skeleton, and skinning as continuous $S^3$ fields over a shared spatial domain. They produce a discrete skeleton through a post-processing clustering step. 
Our approach instead learns explicit surface and skeleton SLats, decodes skeleton topology autoregressively, and predicts skinning through point--bone attention in an end-to-end network. We also introduce open-vocabulary joint labeling for easier motion retargeting.

\section{Method}
\label{sec:method}

\begin{figure}[tb]
  \centering
  \includegraphics[width=\textwidth]{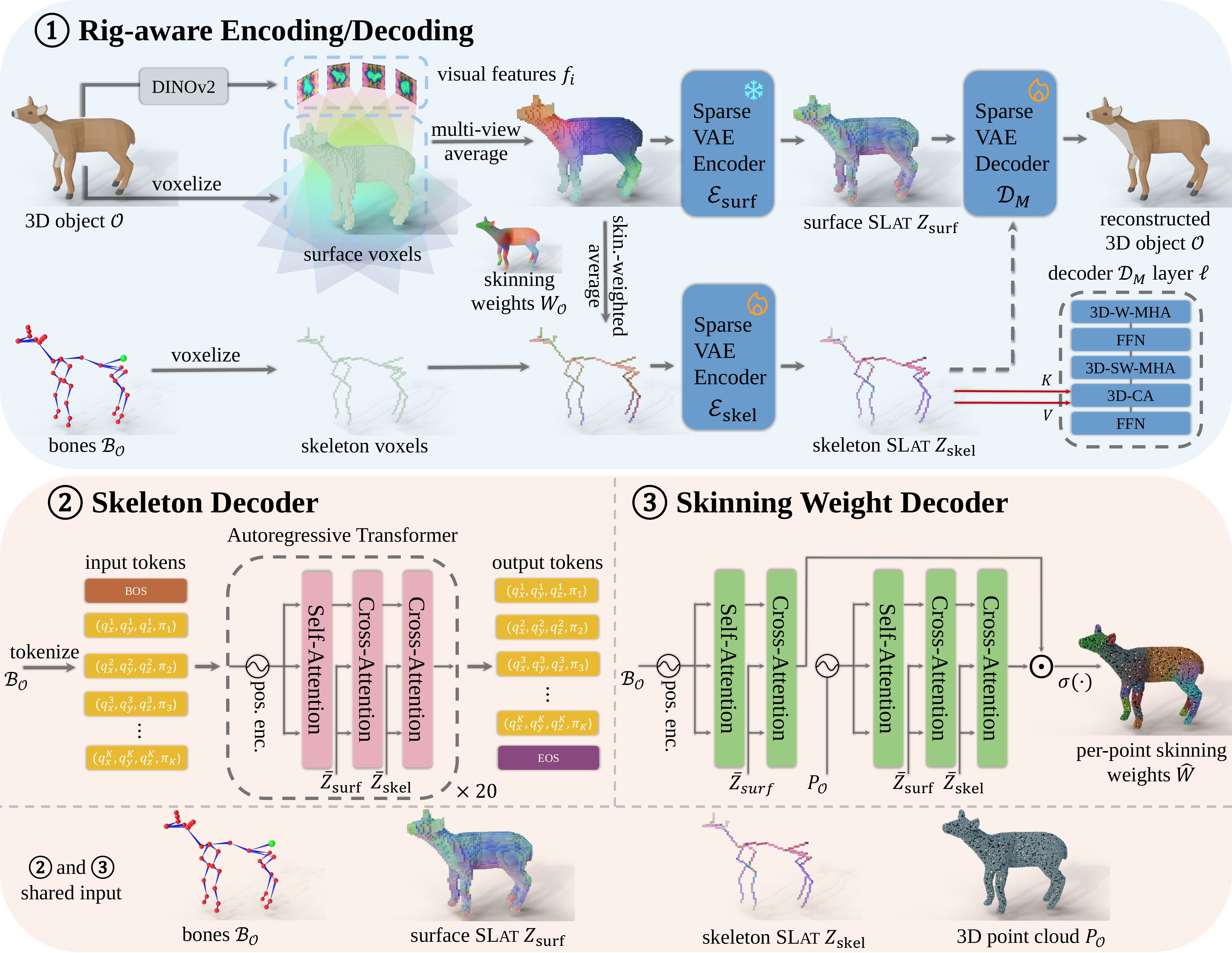}
\vspace{-15pt}   
  \caption{ 
  Overview of the rig-aware autoencoder. 
  A surface encoder produces surface SLats from multiview visual features attached to occupied surface voxels, while a skeleton encoder produces skeleton SLats from rig-aware features attached to voxels intersecting the bones. 
  The two latent representations are jointly decoded into mesh geometry, skeleton structure, and skinning weights.
  }
  \label{fig:autoencoder}
\vspace{-8pt}    
\end{figure}

\begin{figure}[tb]
  \centering
  \includegraphics[width=\textwidth]{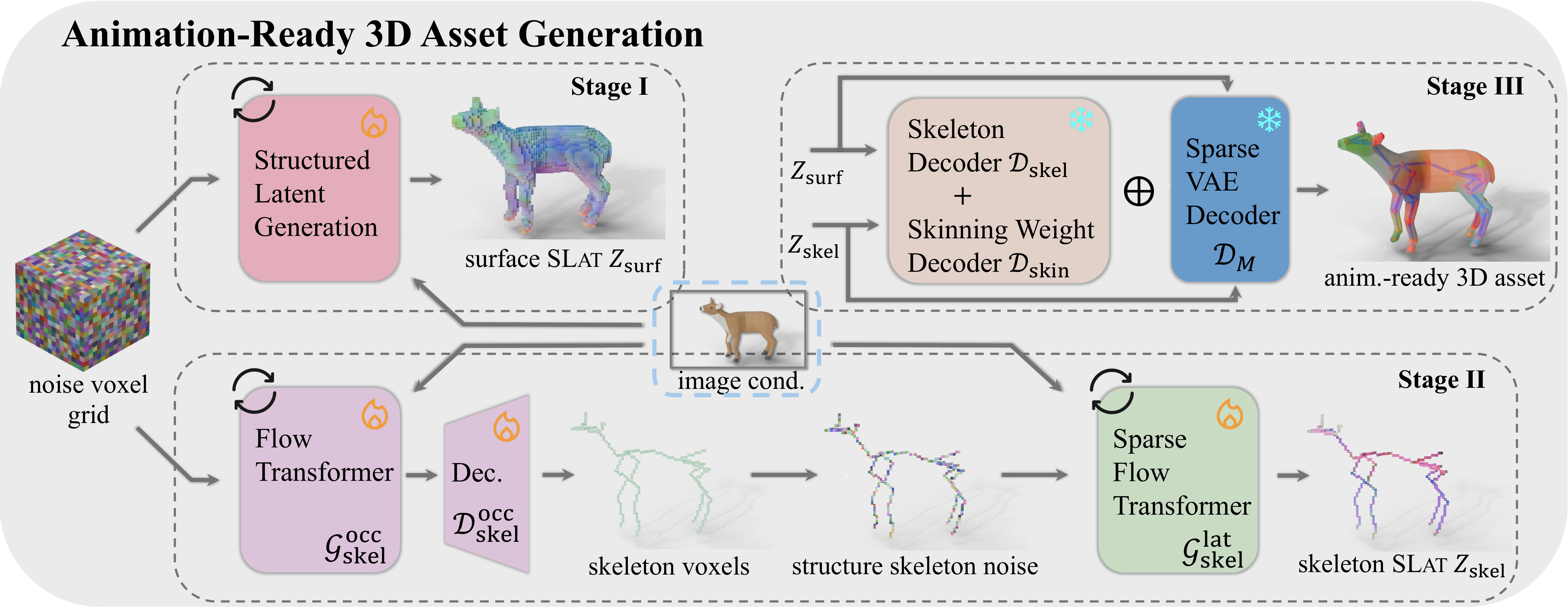}
\vspace{-15pt}      
  \caption{
  Animation-ready asset generation. Our generation pipeline yields both surface and skeleton SLats, which then are decoded into mesh geometry, skeleton structure, and skinning weights.
  }
  \label{fig:generative_model}
\vspace{-8pt}      
\end{figure}

\paragraph{Overview.}
\method{} builds on the Structured LATent representation of TRELLIS~\cite{xiang2024trellis}, which represents 3D assets with sparse voxel locations and local latent codes. 
We extend this representation with two coupled latent sets: surface SLats for geometry and appearance, and skeleton SLats for internal articulation. 
A rig-aware autoencoder (Fig.~ \ref{fig:autoencoder})
maps rigged meshes to these latents and decodes them into mesh geometry, skeleton, and skinning weights; a generative model then synthesizes both latent sets for image-conditioned rigged asset generation (Fig.~\ref{fig:generative_model}).

\subsection{Rig-Aware Autoencoder}
\label{sec:autoencoder}

\paragraph{Surface encoder.}
We adopt the TRELLIS surface encoder~\cite{xiang2024trellis}. 
The input mesh is voxelized at resolution $64^3$, and voxels intersecting the surface are marked active. 
For each active voxel, we aggregate DINOv2 features~\cite{oquab2024dinov2} from multiview renderings, and process the resulting sparse feature grid with a sparse VAE encoder to obtain surface SLats:
\begin{equation}
    \mZ_{\mathrm{surf}} = \{(\vc_i, \vz^{\mathrm{surf}}_i)\}_{i=1}^{L_{\mathrm{surf}}},
\end{equation}
where $\vc_i \in \{0,\ldots,63\}^3$ is an active voxel coordinate and $\vz^{\mathrm{surf}}_i$ is the latent code attached to that voxel. 
These surface SLats provide a local representation of geometry and appearance.

\paragraph{Rig encoder.}
Each training asset contains a skeleton and skinning weights. 
We denote an asset by $\mathcal{O}$, its mesh vertices by $V_{\mathcal{O}}$, and its bones by
$
    \mathcal{B}_{\mathcal{O}} =
    \{(\vj^{\mathrm{head}}_b, \vj^{\mathrm{tail}}_b)\}_{b=1}^{|\mathcal{B}_{\mathcal{O}}|},
$, 
where $\vj^{\mathrm{head}}_b, \vj^{\mathrm{tail}}_b \in \mathbb{R}^3$ are the head and tail joint coordinates of bone $b$. 
The skinning weight matrix is
$
    \mW_{\mathcal{O}} \in \mathbb{R}^{|V_{\mathcal{O}}| \times |\mathcal{B}_{\mathcal{O}}|},
$, where $w_{\mathcal{O}}^{(v,b)}$ is the influence of bone $b$ on vertex $v$.

To attach surface-aware context to the skeleton, we construct a feature for each bone by pooling the multiview DINOv2 features of its influenced vertices. 
Let $\vf_v$ be the multiview feature associated with vertex $v$, obtained by projecting the vertex into the rendered views and averaging the corresponding DINOv2 features. 
For each bone $b$, let
$
    V_b = \{v \in V_{\mathcal{O}} \mid w_{\mathcal{O}}^{(v,b)} > 0\}
$
be the set of vertices influenced by that bone. 
We define the bone feature as the skinning-weighted average:
\begin{equation}
    \vf_b =
    \frac{
        \sum_{v \in V_b} w_{\mathcal{O}}^{(v,b)} \vf_v
    }{
        \sum_{v \in V_b} w_{\mathcal{O}}^{(v,b)} + \epsilon
    },
    \label{eq:bone_feature}
\end{equation}
where $\epsilon$ is a small constant for numerical stability.

We then rasterize each bone segment $(\vj^{\mathrm{head}}_b, \vj^{\mathrm{tail}}_b)$ into the same $64^3$ voxel grid and attach $\vf_b$ to all voxels intersected by the segment, including the voxels containing the head and tail joints. 
If multiple bones intersect the same voxel, their features are averaged. 
Because skeleton voxels are sparse and mostly lie inside the mesh rather than on the surface, we process them with a separate sparse encoder $\mathcal{E}_{\mathrm{skel}}$. 
This produces skeleton structured latents:
\begin{equation}
    \mZ_{\mathrm{skel}} = \{(\vc_i, \vz^{\mathrm{skel}}_i)\}_{i=1}^{L_{\mathrm{skel}}},
\end{equation}
which encode the internal articulation structure of the object. 

\paragraph{Mesh decoder.}
Our mesh decoder extends the TRELLIS mesh decoder~\cite{xiang2024trellis}. 
Given surface SLats $\mZ_{\mathrm{surf}}$, the decoder predicts local FlexiCubes parameters~\cite{shen2023flexicubes} for each active surface voxel and upsamples the sparse representation to a higher resolution before extracting the final mesh. Inspired by structure-aware reconstruction methods that use skeletal or medial information to improve surface reconstruction~\cite{petrov2024gem3d}, we make geometry reconstruction aware of the underlying rig by augmenting each decoder block with a cross-attention layer from surface tokens to skeleton tokens. 
Let $\mH^{\ell}_{\mathrm{surf}}$ be the surface token features at decoder layer $\ell$, and let $\mH_{\mathrm{skel}}$ be the skeleton SLats after projection, positional encoding, and decoder-side processing, used as keys and values in the cross-attention (CA) layer:

\begin{equation}
    \widetilde{\mH}^{\ell}_{\mathrm{surf}}
    =
    \mathrm{CA}
    \left(
        \mH^{\ell}_{\mathrm{surf}},
        \mH_{\mathrm{skel}},
        \mH_{\mathrm{skel}}
    \right),
\end{equation}
followed by the standard TRELLIS decoder operations. This conditions surface reconstruction on the internal articulation structure.
For each active surface voxel, the decoder predicts FlexiCubes parameters and signed distance values,
\begin{equation}
    \mathcal{D}_{M}(\mZ_{\mathrm{surf}}, \mZ_{\mathrm{skel}})
    =
    \{(\vtheta_i, \vd_i)\}_{i=1}^{L_{\mathrm{surf}}},
\end{equation}
where $\vtheta_i$ contains the local FlexiCubes parameters
, and $\vd_i$ denotes SDF values at voxel vertices. 
The final mesh is extracted from the predicted implicit field  following TRELLIS and FlexiCubes.

\paragraph{Skeleton decoder.}
\label{subsec:skeleton_decoder}
We decode the skeleton using an autoregressive transformer conditioned on both surface and skeleton SLats. 
Following recent auto-rigging methods~\cite{song2025puppeteer, Song2025magicarticulate}, we represent a skeleton as a sequence of discrete tokens encoding joint coordinates and parent connectivity. We first convert each skeleton into a breadth-first-search ordering starting from the root joint. 
Joints at the same depth are sorted deterministically by their spatial coordinates, using $z$-$y$-$x$ order. 
Each joint token contains a discretized joint coordinate in a $128^3$ grid and an index pointing to its parent joint. 
We quantize skeleton coordinates at a higher resolution than the SLat voxel grid to reduce joint localization error:
$\vt_k = (\evq_x^k, \evq_y^k, \evq_z^k, \pi_k)$,
where $(\evq_x^k,\evq_y^k,\evq_z^k)$ are discretized coordinates and $\pi_k$ is the joint's parent index. 
For non-root joints, $\pi_k < k$; for the root joint, we set $\pi_k=k$. Unlike prior work that conditions skeleton generation on a single global shape feature, we condition on structured latent representations. 
We first project the surface and skeleton SLats to a common dimension $D_s$ and add positional encodings based on their voxel coordinates:
$
    \bar{\vz}^{\mathrm{surf}}_i
    =
    \phi_{\mathrm{surf}}(\vz^{\mathrm{surf}}_i) + \gamma(\vc_i)$,
    $
    \bar{\vz}^{\mathrm{skel}}_i
    =
    \phi_{\mathrm{skel}}(\vz^{\mathrm{skel}}_i) + \gamma(\vc_i)$, 
where $\phi_{\mathrm{surf}}$ and $\phi_{\mathrm{skel}}$ are learned linear projections and $\gamma$ is a positional encoding function. At transformer layer $\ell$, token features are updated by causal self-attention (CausalSA) followed by cross-attention to the surface and skeleton SLats:
\begin{equation}
    \vh^{\ell}_t
    =
    \mathrm{CausalSA}
    \left(\vh^{\ell-1}_t\right),
    \tilde{\vh}^{\ell}_t
    =
    \mathrm{CA}
    \left(
        \vh^{\ell}_t,
        \bar{\mZ}_{\mathrm{surf}},
        \bar{\mZ}_{\mathrm{surf}}
    \right),
    \vh^{\ell+1}_t
    =
    \mathrm{CA}
    \left(
        \tilde{\vh}^{\ell}_t,
        \bar{\mZ}_{\mathrm{skel}},
        \bar{\mZ}_{\mathrm{skel}}
    \right).
\end{equation}
where $\bar{\mZ}_{\mathrm{surf}}=\{\bar{\vz}^{\mathrm{surf}}_i\}_{i=1}^{L_{\mathrm{surf}}}$ and $\bar{\mZ}_{\mathrm{skel}}=\{\bar{\vz}^{\mathrm{skel}}_i\}_{i=1}^{L_{\mathrm{skel}}}$ are the projected and positionally encoded SLat tokens used as keys and values in the cross-attention layers. 
The transformer outputs categorical distributions over coordinate tokens and valid parent indices. 
During training, the model receives the ground-truth token sequence and is optimized with teacher forcing. 
During inference, generation starts from a BOS token and proceeds  until an EOS token is produced or a maximum joint count is reached. Parent logits are masked so that non-root token $k$ can only select indices $< k$.

\paragraph{Skinning weight decoder.}
\label{subsec:skinning_weight_decoder}
Given a generated or reconstructed mesh and skeleton, we predict skinning weights with an attention-based module. 
During training, the input consists of a fixed-size point cloud sampled from the mesh surface:
$\mP = \{(\vp_i,\vn_i)\}_{i=1}^{N}$,
where $\vp_i \in \mathbb{R}^3$ is a point and $\vn_i \in \mathbb{R}^3$ is its normal, together with the decoded bones $\mathcal{B}_{\mathcal{O}}$ and the two latent representations $\mZ_{\mathrm{surf}}$ and $\mZ_{\mathrm{skel}}$. We embed points and bones into a shared feature dimension $D_p$. 
Each bone is represented by its head and tail coordinates:
$
    \vh^0_{\vp_i}
    =
    \phi_P([\vp_i,\vn_i]) + \gamma(\vp_i),
    \vh^0_{b}
    =
    \phi_B([\vj^{\mathrm{head}}_b,\vj^{\mathrm{tail}}_b])
    + \gamma(\vj^{\mathrm{head}}_b)
    + \gamma(\vj^{\mathrm{tail}}_b).
$

Point and bone tokens are refined through self-attention and cross-attention to surface and skeleton SLats. 

The skinning logits are computed by point--bone similarity, then normalized with softmax:
\begin{equation}
    a_{i,b} =
    \frac{
        \langle \psi_P(\tilde{\vh}_{\vp_i}), \psi_B(\vh_b) \rangle
    }{
        \sqrt{D_p}
    },
    \widehat{w}_{i,b}
    =
    \frac{\exp(a_{i,b})}
    {\sum_{b'=1}^{|\mathcal{B}_{\mathcal{O}}|} \exp(a_{i,b'})}.
    \label{eq:skinning_softmax}    
\end{equation}
where $\psi_P(\tilde{\vh}_{\vp_i})$ and $\psi_B(\vh_b)$ are the output token embeddings corresponding to point $i$ and bone $b$, respectively. At inference time, the same decoder can be evaluated on the vertices of the extracted mesh to obtain per-vertex skinning weights. Since attention and the softmax operate over the decoded bone tokens, the skinning decoder naturally handles skeletons with variable numbers of bones.

\paragraph{Autoencoder training.}
Training jointly supervises mesh reconstruction, skeleton reconstruction, skinning prediction, and latent regularization:
\begin{equation}
    \mathcal{L}_{\mathrm{AE}}
    =
    \mathcal{L}_{\mathrm{mesh}}
    +
    \lambda_{\mathrm{skel}}\mathcal{L}_{\mathrm{skel}}
    +
    \lambda_{\mathrm{skin}}\mathcal{L}_{\mathrm{skin}}
    +
    \lambda_{\mathrm{KL}}\mathcal{L}_{\mathrm{KL}}.
    \label{eq:autoencoder_loss}
\end{equation}
Here $\mathcal{L}_{\mathrm{mesh}}$ is the TRELLIS mesh reconstruction loss, $\mathcal{L}_{\mathrm{skel}}$ is cross-entropy over skeleton coordinate and parent tokens, $\mathcal{L}_{\mathrm{skin}}$ is soft cross-entropy over skinning distributions, and $\mathcal{L}_{\mathrm{KL}}$ regularizes the surface and skeleton latent distributions.

We note that the surface encoder and mesh decoder are initialized from TRELLIS~\cite{xiang2024trellis}, then fine-tuned during the training of our autoencoder. 

\subsection{Animation-Ready Asset Generation}
\label{sec:generation}

We generate rigged assets by learning generative models over the latent representations produced by the rig-aware autoencoder.  We follow the two-stage generation strategy of TRELLIS~\cite{xiang2024trellis}, which first generates the sparse voxel structure and then generates the local latent codes attached to the active voxels. In our setting, we apply this two-stage process to the surface and skeleton SLats.
For each latent type $r \in \{\mathrm{surf},\mathrm{skel}\}$, we represent the active voxel set as a binary occupancy grid $\mO_r \in \{0,1\}^{64 \times 64 \times 64}$. A 3D convolutional VAE compresses this grid into a low-resolution continuous feature grid $\vy_r \in \mathbb{R}^{16 \times 16 \times 16 \times C_r}$.
A rectified-flow transformer $\mathcal{G}^{\mathrm{occ}}_r$ learns to generate $\vy_r$ conditioned on image features extracted from input renders. The VAE decoder then maps $\vy_r$ back to a $64^3$ active voxel grid. 
Given the generated active voxels, a second rectified-flow transformer $\mathcal{G}^{\mathrm{lat}}_r$ generates the latent codes attached to those voxels,
$\mZ_r = \mathcal{G}^{\mathrm{lat}}_r(\mO_r, \mI)$,
where $\mI$ denotes the image condition.  As in TRELLIS, image features are injected into the flow transformers through cross-attention. 
In total, our generative model contains four flow models: occupancy and latent generators for the surface SLats, and occupancy and latent generators for the skeleton SLats.
The latents  $\mZ_{\mathrm{surf}}, \mZ_{\mathrm{skel}}$
are  passed to the mesh decoder, skeleton decoder, and skinning decoder:
\begin{equation}
    \widehat{M}
    =
    \mathcal{D}_M(\mZ_{\mathrm{surf}},\mZ_{\mathrm{skel}}),
    \widehat{\mathcal{B}}
    =
    \mathcal{D}_{\mathrm{skel}}(\mZ_{\mathrm{surf}},\mZ_{\mathrm{skel}}),
    \widehat{\mW}
    =
    \mathcal{D}_{\mathrm{skin}}(\widehat{M},\widehat{\mathcal{B}},\mZ_{\mathrm{surf}},\mZ_{\mathrm{skel}}).
\end{equation}

\section{Open-Vocabulary Joint Label Assignment}
\label{sec:joint_labels}

Many animation pipelines require semantic joint names to establish correspondences between source and target skeletons for motion retargeting. 
However, template-free skeleton generation methods typically output only joint coordinates and connectivity, or generic identifiers such as \texttt{Joint1} and \texttt{Bone003}, which do not encode anatomical, semantic, or functional correspondence. 
A closed-set classifier would restrict generated rigs to a predefined label vocabulary or skeleton template, which is undesirable for assets with non-standard parts, different naming conventions, or object-specific articulations. 
We therefore use an open-vocabulary formulation: generated joints are embedded into a shared vision-language space and can be queried using labels from arbitrary templates.

Given a generated skeleton and the corresponding surface and skeleton SLats, our labeling module predicts a normalized embedding $\ve_{J_k}$ for each joint $k$ in the embedding space of a frozen OpenCLIP text encoder~\cite{ilharco_gabriel_2021_openclip}. 
The module conditions on joint coordinates, skeleton hierarchy, and the global SLat context. 
Since joint labels are highly dependent along the kinematic tree, we use an autoregressive transformer following the BFS ordering of the skeleton decoder: each joint embedding attends to the surface and skeleton SLats, causally attends to previously generated joints, and cross-attends to embeddings of previous labels. 
This autoregressive variant performs best in our experiments and helps disambiguate repeated or symmetric structures such as left/right limbs, fingers, fins, and tails. We train the joint embedding model using cleaned joint labels from Anymate~\cite{deng2025anymate}. 
Because artist-provided labels are heterogeneous and often contain armature prefixes, namespaces, duplicated identifiers, or uninformative strings, we preprocess them with an LLM-based cleanup pipeline using \textsc{Qwen3-8B}~\cite{yang2025qwen3technicalreport}; details and examples are provided in the appendix. 
For each cleaned label $\ell_k$, we compute a normalized text embedding $\ve_{\ell_k}=\mathrm{CLIP}_{\mathrm{text}}(\ell_k)$ and optimize an InfoNCE objective over joint-label pairs:
\begin{equation}
    \mathcal{L}_{\mathrm{label}}
    =
    -\frac{1}{B}
    \sum_{k=1}^{B}
    \log
    \frac{
        \exp(\ve_{J_k}^{\top} \ve_{\ell_k}/\tau)
    }{
        \sum_{m=1}^{B}
        \exp(\ve_{J_k}^{\top} \ve_{\ell_m}/\tau)
    },
    \label{eq:label_contrastive}
\end{equation}
where $\tau$ is a temperature. Other labels in the minibatch serve as negatives. 
At inference time, each generated joint is labeled by nearest-neighbor retrieval over any candidate vocabulary, including joint names of a retargeting template. 
Thus, the learned joint representation is decoupled from the choice of label set, allowing the same rig to be matched to different animation templates without retraining.

\begin{figure}[t]
    \centering
    \includegraphics[
        width=1.000\textwidth,
        height=1.000\textheight,
        keepaspectratio
    ]{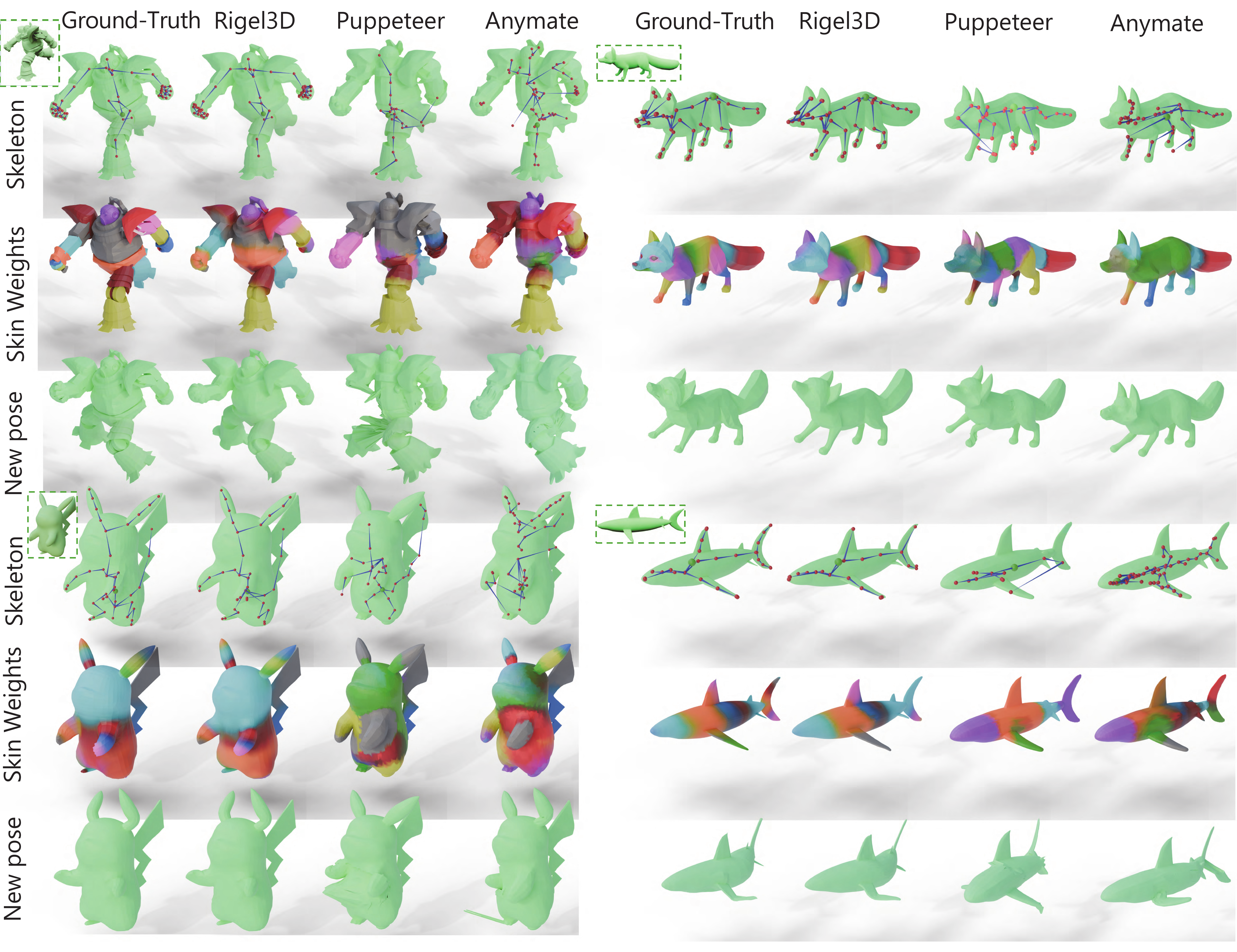}
    \vspace{-20pt}
    \caption{\textbf{Qualitative comparison with Anymate and Puppeteer.} 
    Compared to auto-rigging baselines, \method{} produces rigs that better match the reference ones in joint placement, connectivity, and skinning while yielding more coherent novel poses. Shapes are shown without texture to emphasize skeleton and pose geometry rather than appearance.
    Skinning colors indicate bone influence, with smooth transitions corresponding to blended weights. Green insets show input images.
    }  \label{fig:comparison_with_puppeteer_anymate}
\vspace{-12pt}        
\end{figure}

\begin{figure}[t]
    \centering
    \includegraphics[
        width=1.000\textwidth,
        height=1.000\textheight,
        keepaspectratio
    ]{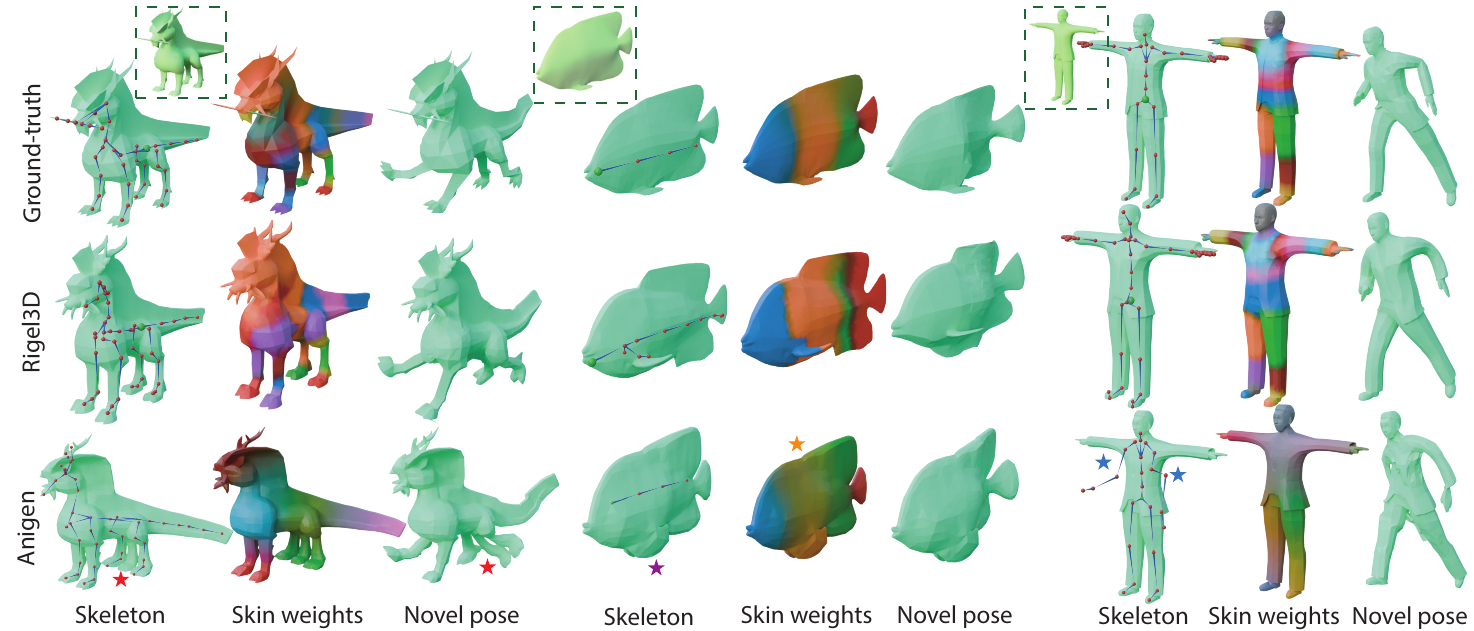}
    \vspace{-15pt}
    \caption{\textbf{Qualitative comparison with AniGen.}
    Here we compare generated shapes, skeletons, skinning, and novel-pose deformations. Green insets show input images.
    \method{} better preserves the articulation structure and produces more coherent skinning weights across diverse assets. 
    Stars highlight issues: \textcolor{red}{$\star$} an extra generated spurious limb, \textcolor{purple}{$\star$} incomplete fin geometry, \textcolor{orange}{$\star$} skinning weights that differ substantially from the reference distribution, and \textcolor{blue}{$\star$} erroneous joint placement.}
    \label{fig:comparison_with_anigen}
    \vspace{-12pt}    
\end{figure}

\section{Experiments}
\label{sec:experiments}
\label{sec:results}

We now discuss our training and evaluation protocol, along with comparisons, results, and ablations.

\paragraph{Training dataset.}
We train \method{} on the Anymate dataset~\cite{deng2025anymate}, the largest publicly available dataset of rigged 3D assets. 
We use its training split, which contains approximately $225$K rigged models with mesh geometry, skeletons, and skinning weights.
To prepare the images used for the extraction of the multi-view Dino features as well as the rig-features, we follow the rendering procedure of TRELLIS \cite{xiang2024trellis}.

\paragraph{Competing methods.}
We compare against two representative auto-rigging baselines, Anymate~\cite{deng2025anymate} and Puppeteer~\cite{song2025puppeteer}, as well as the \emph{concurrent} image-to-rigged-asset generation method AniGen~\cite{huang2026anigen}. 
Anymate and Puppeteer assume an input mesh rather than generating one. 
For a fair comparison in the image-conditioned setting, we provide both methods with meshes generated by TRELLIS~\cite{xiang2024trellis} finetuned on the Anymate training split, matching the training data used by \method{}. 
For Anymate, we use the publicly released checkpoints trained on the same dataset as ours. 
For Puppeteer, we train the model from scratch on our training data, since the released checkpoints are trained on different datasets. 
For AniGen, we use the publicly available checkpoints provided by the authors.

\paragraph{Evaluation protocol.}
We evaluate on the Anymate test split, containing approximately $5.6$K assets~\cite{deng2025anymate}, and on the ModelsResource test split~\cite{xu2020rignet}, containing $270$ diverse rigged assets. 
For each asset, we render $4$ views and use the rendered images as conditioning inputs. 
Our rendering setup and camera parameters follow TRELLIS~\cite{xiang2024trellis}. 
We report metrics averaged over all rendered views and all test assets. 
For comparisons with AniGen~\cite{huang2026anigen}, we evaluate on the ArticulationXL test split used in their work, and the ModelsResource test split. 
We note that our ArticulationXL numbers differ from those reported by AniGen because, at the time of submission, their test images, rendering parameters, and evaluation code were not publicly available. 
We thus evaluate AniGen checkpoints using our rendering setup and  protocol.

\paragraph{Evaluation metrics.}
We use the standard skeleton and skinning metrics introduced in RigNet~\cite{xu2020rignet} and Anymate~\cite{deng2025anymate}. 
For skeleton evaluation, \emph{J2J} measures the symmetric Chamfer distance between predicted and reference joint sets. 
\emph{J2B} measures the average of two distances: from predicted joints to the nearest point on the reference bones, and from reference joints to the nearest point on the predicted bones. 
\emph{B2B} measures the Chamfer distance between predicted and reference bones, treated as line segments. Note that all geometric distances are computed after normalizing each test asset to a unit bounding box.
For skinning evaluation, we report the per-vertex $\ell_1$ and $\ell_2$ distances between predicted and reference skinning weight vectors. In addition, we report the the $KL$ divergence measuring how much the predicted skinning weight distribution differs from the reference skinning weight distribution. 
Since the generated meshes do not share vertex correspondence with the reference mesh, we follow Anigen ~\cite{huang2026anigen} and align predicted skinning weights to the reference skeleton using optimal transport induced by the Wasserstein distance between predicted and reference joints. We report per-point $\ell_1$, $\ell_2$, and KL divergence between the aligned predicted skinning distributions and the reference skinning weights. This avoids assuming a fixed joint ordering and amount or shared mesh topology across methods.

\paragraph{Comparisons with prior auto-rigging methods.}
Table~\ref{tab:img_2_skeleton_metrics} compares \method{} against Anymate and Puppeteer on the Anymate and ModelsResource test splits. 
Both baselines are post-hoc auto-rigging methods that require an input mesh; in this image-conditioned setting, we provide them with meshes generated by the same TRELLIS model used in our pipeline. 
On Anymate, \method{} outperforms both baselines across all reported skeleton and skinning metrics. 
On ModelsResource, \method{} obtains the best score on J$2$J, B$2$B, and all skinning metrics, while Anymate is slightly better on J$2$B. 
The gains are most pronounced for B$2$B, suggesting that our skeletons better capture bone-level structure and connectivity, and for KL, indicating improved alignment of the predicted skinning distributions. 
Overall, these results support our design choice of jointly modeling surface geometry and rig structure, rather than applying rigging as a post-processing step to a generated mesh.
Qualitative comparisons in Figure~\ref{fig:comparison_with_puppeteer_anymate} further illustrate that \method{} produces more accurate joint placement, bone connectivity, and skinning weight distributions on complex shapes.

\begin{table}[t!]
  \centering
  \caption{Image-conditioned rig generation comparisons (skeleton distances are reported $\times100$).}
  \vspace{-7pt}
  \resizebox{\textwidth}{!}{
  \begin{tabular}{@{}l cccccc cccccc@{}}
    \toprule
     & \multicolumn{6}{c}{Anymate} & \multicolumn{6}{c}{ModelsResource} \\
    \cmidrule(lr){2-7} \cmidrule(lr){8-13}
     & J$2$J$\downarrow$ & J$2$B$\downarrow$ & B$2$B$\downarrow$ &$\ell_1$$\downarrow$ & $\ell_2$$\downarrow$ & KL$\downarrow$ & J$2$J$\downarrow$ & J$2$B$\downarrow$ & B$2$B$\downarrow$ & $\ell_1$$\downarrow$ & $\ell_2$$\downarrow$ & KL$\downarrow$ \\
    \midrule
    Anymate & $7.834$ & $5.964$ & $5.363$ & $0.0449$ & $0.0180$ & $2.589$ & $5.834$ & $\mathbf{4.184}$ & $4.322$ & $0.0661$ & $0.0285$ & $2.215$ \\ %
    Puppeteer & $8.645$ & $6.739$ & $5.710$ & $0.0457$ & $0.0183$ & $2.611$ & $6.797$ & $5.049$ & $4.808$ & $0.0673$ & $0.0294$ & $2.249$ \\ %
    \textbf{Ours} & $\mathbf{7.364}$ & $\mathbf{5.736}$ & $\mathbf{4.902}$ & $\mathbf{0.0440}$ & $\mathbf{0.0176}$ & $\mathbf{2.478}$ & $\mathbf{5.765}$ & $4.260$ & $\mathbf{4.200}$ & $\mathbf{0.0654}$ & $\mathbf{0.0283}$ & $\mathbf{2.155}$ \\ %
    \bottomrule
  \end{tabular}
  }
  \label{tab:img_2_skeleton_metrics}     
  \vspace{-5pt}
\end{table}

\paragraph{Quantitative comparisons with AniGen.}
Table~\ref{tab:img2skel_vs_anigen} compares \method{} with the concurrent image-to-rigged-asset method AniGen~\cite{huang2026anigen} on ArticulationXL and ModelsResource. 
On ArticulationXL, \method{} substantially improves all skeleton metrics over AniGen, reducing J2J from $10.667$ to $7.168$ ($32.8\%$ relative reduction), J2B from $9.715$ to $6.036$ ($37.9\%$), and B2B from $8.601$ to $5.976$ ($30.5\%$).
AniGen obtains slightly lower skinning errors on this split, suggesting that its continuous field representation can produce competitive skinning distributions once the skeleton is aligned. 
On ModelsResource, \method{} outperforms AniGen across all reported skeleton and skinning metrics, with especially large gains in skeleton accuracy. 
These results indicate that explicitly modeling skeleton structure with skeleton SLats and autoregressive topology decoding improves the quality of generated rigs, while remaining competitive on skinning prediction. 
Qualitative comparisons in Fig.~\ref{fig:comparison_with_anigen} further illustrates that \method{} better preserves shape structure, places joints more consistently, and produces skinning weights that more closely match the reference distribution.
\begin{table}[t]
  \centering
  \caption{Comparisons with Anigen (skeleton distances are reported $\times100$).}
  \vspace{-7pt}
  \resizebox{\textwidth}{!}{
  \begin{tabular}{@{}l cccccc cccccc@{}}
    \toprule
     & \multicolumn{6}{c}{Articulation XL} & \multicolumn{6}{c}{ModelsResource} \\
    \cmidrule(lr){2-7} \cmidrule(lr){8-13}
     & J$2$J$\downarrow$ & J$2$B$\downarrow$ & B$2$B$\downarrow$ & $\ell_1$$\downarrow$ & $\ell_2$$\downarrow$ & KL$\downarrow$ & J$2$J$\downarrow$ & J$2$B$\downarrow$ & B$2$B$\downarrow$ &  $\ell_1$$\downarrow$ & $\ell_2$$\downarrow$ & KL$\downarrow$ \\
    \midrule
    AniGen & $10.667$ & $9.715$ & $8.601$ & $\mathbf{0.0609}$ & $\mathbf{0.0249}$ & $\mathbf{2.293}$ & $9.730$ & $7.740$ & $6.852$ & $0.0669$ & $0.0293$ & $2.330$ \\
    \textbf{Ours} & $\mathbf{7.168}$ & $\mathbf{6.036}$ & $\mathbf{5.976}$ & $0.0637$ & $0.0267$ & $2.320$ & $\mathbf{5.765}$ & $\mathbf{4.260}$ & $\mathbf{4.200}$ & $\mathbf{0.0654}$ & $\mathbf{0.0283}$ & $\mathbf{2.155}$ \\
    \bottomrule
  \end{tabular}
  }
  \label{tab:img2skel_vs_anigen}  
  \vspace{-12pt}
\end{table}

\paragraph{Joint labeling.}
Fig.~\ref{fig:teaser} illustrates open-vocabulary joint labeling on representative generated characters. 
The predicted labels identify semantically meaningful joints across different body structures, enabling correspondences to downstream motion-retargeting templates without assuming a fixed skeleton vocabulary (please see the supplement for motion retargeting examples). 
We provide a detailed comparison of alternative joint-labeling strategies in the Appendix.

\paragraph{Ablation studies and additional comparisons.}
We provide ablation studies in the Appendix analyzing the contribution of our main architectural choices, including the benefit of factorizing the latent space into distinct surface and skeleton SLats and coupling them during decoding.
We also compare the shape generation quality of \method{} against TRELLIS~\cite{xiang2024trellis}.

\section{Discussion and Limitations}
\label{sec:conclusion}

We introduced \method{}, a generative framework for producing animation-ready 3D assets with geometry, skeletons, skinning weights, and joint labels. 
By jointly modeling surface and skeleton SLats, our method narrows the gap between static 3D generation and rig-based animation workflows.

\paragraph{Limitations.}
Generated rigs may still contain missing joints, spurious branches, or imperfect connectivity. 
Because skinning is not supervised by long motion sequences, deformations can be less natural under extreme poses or specific animation intents. 
Open-vocabulary labels may be ambiguous for repeated structures such as fingers, tails, or decorative appendages. Finally, our method inherits limitations of image-conditioned 3D generation when input views are ambiguous or occluded.

\section*{Acknowledgements}
This project has received funding from the European Research Council (ERC) under the Horizon Research and 
Innovation Programme (Grant agreement No. 101124742).

\bibliographystyle{plainnat}
\bibliography{main}
\newpage

\appendix

\section{Detailed description of Open-Vocabulary Joint Label Assignment}
\label{sec:joint_labels_appendix}

Many downstream animation pipelines assume that the joints of an input skeleton are labeled. 
For example, motion retargeting methods often require correspondences between source and target joints, such as matching \texttt{LeftArm}, \texttt{Spine}, or \texttt{Head} across different characters. 
However, template-free skeleton generation methods typically output only joint coordinates and connectivity, without semantic joint names. 
This limits their compatibility with existing animation tools and retargeting pipelines, where joint labels provide the semantic bridge between different skeletons.

A closed-set joint classifier would restrict retargeting to a predefined label vocabulary or skeleton template, which is undesirable for generated assets whose rigs may contain non-standard parts, different naming conventions, or object-specific articulations. 
Instead, we use an open-vocabulary formulation so that joints can be queried using labels from arbitrary source templates, including human, animal, creature, and object rigs. 
This enables retargeting pipelines to establish correspondences to a wider range of templates without retraining the labeler for each label set.

We therefore introduce an open-vocabulary joint labeling module as a post-processing component for generated rigs. 
Given a generated skeleton and the corresponding surface and skeleton SLats, the module embeds each joint into a shared vision-language space and retrieves semantic labels using text queries. 
This design allows the model to assign labels beyond a fixed closed vocabulary, while still exploiting the geometry, hierarchy, and rig-aware latent context produced by our method.
Traditional artist-created rigs usually include manually assigned joint or bone labels.  
These labels often encode the body part or object part associated with the joint, such as \texttt{Torso}, \texttt{Arm}, \texttt{Tail}, or \texttt{Handle}, as well as relative location descriptors such as \texttt{Left}, \texttt{Right}, \texttt{Upper}, or \texttt{Lower}. 
In contrast, prior template-free rigging methods~\cite{xu2020rignet, Song2025magicarticulate, song2025puppeteer, liu2025riganything, deng2025anymate, sun2025armo, zhang2025unirig, guo2025autoconnectconnectivitypreservingrigformerdirect} typically generate structural skeletons only. 
Their output joints are either unlabeled or assigned generic identifiers such as \texttt{Joint1} or \texttt{Bone003}, which do not encode anatomical, semantic, or functional correspondences needed for retargeting.

\paragraph{Label preprocessing.}
The Anymate dataset~\cite{deng2025anymate} provides text labels for many joints, but these labels are collected from heterogeneous artist-created assets and are not standardized. 
They may contain armature prefixes, namespace strings, duplicated identifiers, inconsistent left/right conventions, or non-informative names. 
We clean these labels using an LLM-based preprocessing pipeline with \textsc{Qwen3-8B}~\cite{yang2025qwen3technicalreport}.  
The preprocessing removes asset-specific prefixes and suffixes, normalizes common side and part descriptors, and filters samples whose labels are uninformative or ambiguous. 
Examples are shown in Tab.~\ref{tab:labels_before_after}; the full prompt and filtering rules are provided in Section \ref{sec:label_data_details}.

\paragraph{Joint-text embedding model.}
We train a model that maps each joint to the embedding space of a frozen OpenCLIP text encoder~\cite{ilharco_gabriel_2021_openclip}. 
Let $\vj_k \in \mathbb{R}^3$ denote the coordinate of joint $k$. 
The model receives the joint coordinates, the skeleton hierarchy, and the surface and skeleton SLats. 
We first embed joints with positional encodings:
\begin{equation}
    \vh^0_{\vj_k} = \phi_J(\vj_k) + \gamma(\vj_k).
\end{equation}
The joint embeddings are processed by transformer blocks that combine self-attention over joints with cross-attention to the structured latent representations:
\begin{align}
    \mH'_J
    &=
    \mathrm{SelfAttn}(\mH^0_J),
    \\
    \mH''_J
    &=
    \mathrm{CrossAttn}
    \left(
        \mH'_J,
        \bar{\mZ}_{\mathrm{surf}},
        \bar{\mZ}_{\mathrm{surf}}
    \right),
    \\
    \mH'''_J
    &=
    \mathrm{CrossAttn}
    \left(
        \mH''_J,
        \bar{\mZ}_{\mathrm{skel}},
        \bar{\mZ}_{\mathrm{skel}}
    \right),
    \\
    \ve_J
    &=
    \mathrm{MLP}(\mH'''_J).
\end{align}
The output $\ve_J$ is projected to the CLIP text embedding dimension and $\ell_2$-normalized.
For each cleaned joint label $\ell_k$, we compute a normalized text embedding:
\begin{equation}
    \ve_{\ell_k} = \mathrm{CLIP}_{\mathrm{text}}(\ell_k).
\end{equation}

We train the joint embedding model with an InfoNCE contrastive objective over joints and labels in a minibatch:
\begin{equation}
    \mathcal{L}_{\mathrm{label}}
    =
    -\frac{1}{B}
    \sum_{k=1}^{B}
    \log
    \frac{
        \exp(\ve_{J_k}^{\top} \ve_{\ell_k}/\tau)
    }{
        \sum_{m=1}^{B}
        \exp(\ve_{J_k}^{\top} \ve_{\ell_m}/\tau)
    },
    \label{eq:label_contrastive_appendix}
\end{equation}
where $\tau$ is a learned or fixed temperature. Labels from other joints in the minibatch serve as negatives. 
At inference time, each generated joint can be labeled by nearest-neighbor retrieval among any candidate vocabulary, including the joint names of a target retargeting template, or queried directly with arbitrary text prompts such as \texttt{left wrist}, \texttt{tail base}, or \texttt{front wheel}. This formulation separates the learned joint representation from the choice of label vocabulary, allowing the same generated rig to be matched against different retargeting templates at inference time. Note that because left--right and repeated structures can be ambiguous from local geometry alone, the model conditions on both the full skeleton  and the global SLat context when predicting joint embeddings.

\paragraph{Autoregressive labeling variant.}
We also evaluate an autoregressive variant that predicts joint embeddings while conditioning on previously labeled joints. 
This is useful because joint labels are not independent: for example, the presence of \texttt{LeftUpperArm} increases the likelihood of nearby descendants such as \texttt{LeftForeArm} and \texttt{LeftHand}. 
Following the BFS ordering used by the skeleton decoder, each transformer block performs cross-attention to the latent grids, causal self-attention over previous joints, and causal cross-attention to the embeddings of previous labels:
\begin{align}
    \mH'_J
    &=
    \mathrm{CrossAttn}
    \left(
        \mH^0_J,
        \bar{\mZ}_{\mathrm{surf}},
        \bar{\mZ}_{\mathrm{surf}}
    \right),
    \\
    \mH''_J
    &=
    \mathrm{CrossAttn}
    \left(
        \mH'_J,
        \bar{\mZ}_{\mathrm{skel}},
        \bar{\mZ}_{\mathrm{skel}}
    \right),
    \\
    \mH'''_J
    &=
    \mathrm{CausalSelfAttn}(\mH''_J),
    \\
    \mH^*_J
    &=
    \mathrm{CausalCrossAttn}(\mH'''_J, \mE_T, \mE_T),
    \\
    \ve_J
    &=
    \mathrm{MLP}(\mH^*_J),
\end{align}
where $\mE_T$ denotes the sequence of CLIP embeddings of previous labels, using ground-truth labels during training and generated labels during inference. 
The model is trained with the same contrastive objective as Eq.~\eqref{eq:label_contrastive}, using teacher forcing over the ground-truth label sequence.  

\begin{figure}[tb]
  \centering
  \includegraphics[width=\textwidth]{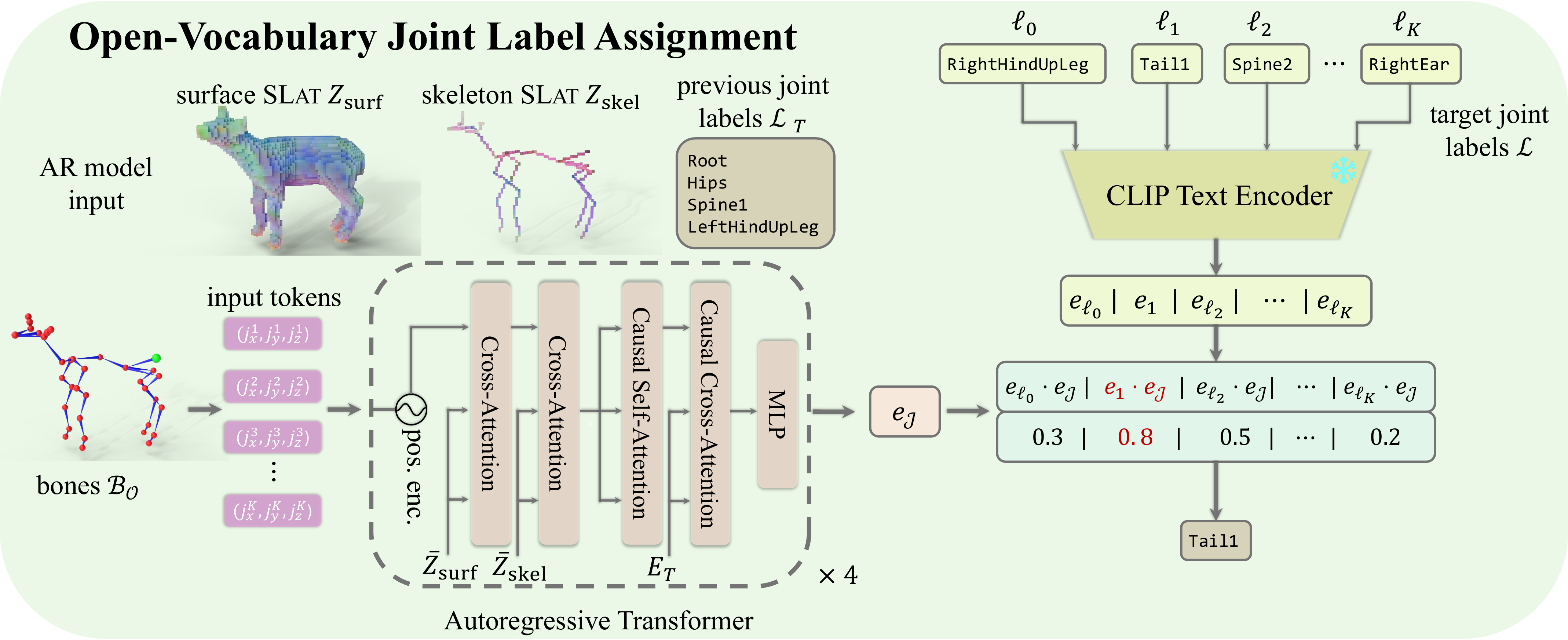}
\vspace{-15pt}      
  \caption{
  Open-vocabulary joint label assignment. Our model coembeds skeleton geometry and text-captions, allowing for label querying and further compatibility with downstream animation pipelines.
  }
  \label{fig:joint_labeling}
\vspace{-8pt}      
\end{figure}

\section{Training and Implementation Details}
\label{sec:training_details}
In this section we provide details regarding hyperparameters and the training and hardware setups.

We finetune $\mathcal{E}_{surf}$ and $\mathcal{D}_M$ on the Anymate \cite{deng2025anymate} training split for $1$ day on $4$ RTX $6000$ Ada $48$GB GPUs starting from the provided TRELLIS \cite{xiang2024trellis} checkpoints and retaining the same loss weighting parameters.
The joint training procedure of $\mathcal{E}_{skel}$, $\mathcal{D}_M$ (including the extra cross-attention layer), skeleton generation and skinning models is done under the same hardware setup.
We randomly sample $5$k assets from the Anymate training split to use as a validation set and train for $\sim 10$ days until convergence, with a batch size of $4$ per GPU.
We set $\lambda_{skel}=1$ and $\lambda_{skin}=1$.
As mentioned in the main text, due to memory constraints we freeze $\mathcal{E}_{surf}$ and pre-extract the surface SLats for each item in the training split.

We train the surface and skeleton Sparse Structure VAEs for $24$ hours on $4$ NVIDIA A$5000$ $24$GB GPUs with a batch size of $4$ per GPU.
We finetune $\mathcal{G}^{occ}_{skel},\mathcal{G}^{occ}_{surf},\mathcal{G}^{lat}_{skel},\mathcal{G}^{lat}_{surf}$ for $2$ days each on $8$ NVIDIA A$5000$s with a batch size of $2$ per GPU, this time starting from the pretrained TRELLIS checkpoints. We found that the provided TRELLIS checkpoints slightly improve performance compared to training from scratch even for the skeleton flow models.

The skeleton labeling  transformer is trained on $4$ NVIDIA A$5000$s GPUs with a batch size of $16$ per GPU for the AR variant and $32$ per GPU for the regular variant.
Training takes $\sim 3$ days until convergence, with a randomly sampled validation set of $5$k items.

For all training procedures, we use the AdamW optimizer \cite{loshchilov2019adamW} with a learning rate of $1\times 10^{-4}$.
For the joint training of the rig-aware autoencoder, skeleton and skinning models specifically, we use a cosine learning rate scheduler with $5000$ warmup steps.

All parameters regarding the noise scheduler and samplers of the $4$ latent flow models are the same as \cite{xiang2024trellis}.
For all experiments we use $25$ sampling steps and set the classifier-free guidance weight to $5.0$.

\section{Additional Comparisons and Ablation Studies}

\paragraph{Comparison with Trellis.} To assess geometric reconstruction quality,  generation quality and the effect of the skeleton-aware latent space and joint training scheme, we evaluate our method against TRELLIS finetuned on the Anymate dataset.
We compare on a selection of metrics curated from \cite{xiang2024trellis} to quantitatively measure the quality of the results.
Table \ref{tab:geom_results} contains quantitative results for geometry reconstruction and image-to-3D generation.
PSNR, LPIPS, Chamfer Distance and F-score measure the reconstruction fidelity of our rig-aware autoencoder.
The first two are computed between renders of the reconstructed and ground truth meshes, whereas for CD and F-score we unproject the depth maps of $100$ renders from uniformly sampled views into a pointcloud and sample $100$k points for computing the metrics.
For FD$_{dinov2}$ and KD$_{dinov2}$ we render $4$ images of the GT and generated asset with yaw angles every $90^{\circ}$ and a pitch of $30^{\circ}$.
For CLIP similarity we use $8$ images rendered with yaw at every $45^{\circ}$ and the same pitch used previously.
\method{} remains competitive with TRELLIS on geometric reconstruction and image-to-3D quality despite jointly training for geometry, skeletons, and skinning. 
While TRELLIS is slightly better on LPIPS, CD, and CLIP, our method matches its F-score, slightly improves PSNR, and yields significantly better distributional feature metrics, reducing FD$_{\text{DINOv2}}$ from $34.14$ to $30.83$ ($9.7\%$ relative reduction) and KD$_{\text{DINOv2}}$ from $12.30$ to $9.35$ ($24.0\%$ relative reduction).

\begin{table}[tb]
  \caption{Geometric reconstruction and Image-to-$3$D comparisons with TRELLIS}
  \label{tab:geom_results}
  \centering
  \begin{tabular}{@{}lcccc | ccc@{}}
    \toprule
     & PSNR$\uparrow$ & LPIPS$\downarrow$ & CD$\downarrow$ & F-score$\uparrow$ & CLIP$\uparrow$ & FD$_{\text{dinov2}}$$\downarrow$ & KD$_{\text{dinov2}}$$\downarrow$ \\
    \midrule
    TRELLIS & $31.077$ & $\mathbf{0.0373}$ & $\mathbf{0.3677}$ & $98.64$ & $\mathbf{89.30}$ & $34.14$ & $12.30$ \\
    \textbf{Ours} & $\mathbf{31.130}$ & $0.0382$ & $0.3704$ & $98.64$ & $88.73$ & $\mathbf{30.83}$ & $\mathbf{9.35}$ \\
  \bottomrule
  \end{tabular}
\end{table}

\paragraph{Ablation study.} 
We conduct an ablation study on some of the architecture choices for our skeleton generation and skinning models.
Since training our method requires access to the ground-truth rigging information, it is unfair to evaluate on the standard geometry conditioned (mesh input) autorigging setting where such information is unavailable during test time.
However, because the standard setting is faster to train and evaluate since it does not require the additional training of the flow models or averaging across multiple views, we utilize this setting to perform an ablation study.
Table \ref{tab:our_ablation_skel_results} shows the mesh-conditioned skeleton generation results for the ablation variants of our method. \emph{Ours (Queries)} refers to a variant of our method that compresses the variable length surface SLats into a fixed-sized shape condition for the autoregressive skeleton generator via a learnable Query set, similarly to how Michelangelo \cite{zhao2023michelangelo} is used to obtain fixed-size global shape information in \cite{Song2025magicarticulate, song2025puppeteer}.
\emph{Ours (joint SLat grid)} refers to our method with the encoder input consisting of both the surface DinoV2 features as well as the rig-aware features within the \emph{same} 3D grid.
Finally, in \emph{Ours (Skeleton SLats)} we perform a single cross-attention operation with just the rig slats inside each transformer block.
Our method, which uses cross-attention with both the surface and skeleton SLats, showcases the effectiveness of the coarse-to-fine, shape-to-skeleton attention based block.
For skinning we include the results for the combined SLat grid.
\emph{Separating the two SLat types for the cross-attention operations significantly improves the model performance.}

\begin{table}[tb]
  \caption{
  Ablation of our method for Mesh-to-Skeleton generation results (values reported $\times100$).
  }
  \label{tab:our_ablation_skel_results}
  \centering
  \begin{tabular}{@{}l ccc ccc@{}}
    \toprule
     & \multicolumn{3}{c}{Anymate} & \multicolumn{3}{c}{ModelsResource} \\
    \cmidrule(lr){2-4} \cmidrule(lr){5-7}
     & J$2$J$\downarrow$ & J$2$B$\downarrow$ & B$2$B$\downarrow$ & J$2$J$\downarrow$ & J$2$B$\downarrow$ & B$2$B$\downarrow$ \\
    \midrule
    \textbf{Ours (Queries)} & $3.826$ & $2.787$ & $2.837$ & $4.105$ & $2.835$ & $3.165$ \\
    \textbf{Ours (Joint SLat grid)} & $3.471$ & $2.472$ & $2.672$ & $3.873$ & $2.587$ & $3.010$ \\
    \textbf{Ours (Skeleton SLats)} & $1.431$ & $1.098$ & $1.909$ & $1.322$ & $\mathbf1.038$ & $2.111$ \\
    \textbf{Ours} & $\mathbf{1.386}$ & $\mathbf{1.068}$ & $\mathbf{1.881}$ & $\mathbf{1.280}$ & $\mathbf{1.006}$ & $\mathbf{2.052}$ \\
    \bottomrule
  \end{tabular}
\end{table}

\begin{table}[tb]
  \caption{
    Ablation study for the skinning weight model, evaluated on the standard mesh-conditioned setting (values reported $\times100$). 
  }
  \label{tab:skinning_weight_metrics}
  \centering
  \begin{tabular}{@{}l ccc ccc@{}}
    \toprule
     & \multicolumn{3}{c}{Anymate} & \multicolumn{3}{c}{ModelsResource} \\
    \cmidrule(lr){2-4} \cmidrule(lr){5-7}
     & L$1$$\downarrow$ & Precision$\uparrow$ & Recall$\uparrow$ & L$1$$\downarrow$ & Precision$\downarrow$ & Recall$\downarrow$ \\
    \midrule
    Ours (Joint SLat grid) & $1.251$ & $83.48$ & $64.85$ & $1.772$ & $78.25$ & $80.89$ \\
    \textbf{Ours} & $1.013$ & $88.19$ & $65.19$ & $1.280$ & $85.66$ & $83.02$ \\
    \bottomrule
  \end{tabular}
\end{table}

\paragraph{Joint Labeling.}

We evaluate the accuracy of our labeling model on the Anymate test-split.
Since there is no relevant method for this task, we compare our model against the non-autoregressive baseline, and report ablations on the same table. 
Table \ref{tab:joint_clip_results} reports the results of our Open-vocabulary joint labeling method on label assignment accuracy.  
\emph{Ours (TRELLIS SLats)} refers to the non-autoregressive variant described in Section \ref{sec:joint_labels}, using only the surface SLats obtained from the finetuned TRELLIS encoder.
\emph{Ours (both SLats)} adds an additonal cross-attention layer with our rig-aware SLats, showcasing the importance of skeleton spatial information and the effectiveness of our latent space.
Finally, \emph{Ours (AR X)} reports the performance of the autoregressive model that also utilizes the CLIP embeddings of the previously labeled joints. 
We experiment with various traversals of the skeleton tree.
"Default order" refers to no explicit reordering of the joint sequence (i.e. we retain the parent-child order hierarchy of the artist) whereas the other variants are standard tree traversals and spatial ordering.

\label{subsec:joint_clip_results}
\begin{table}[tb]
  \caption{Joint label assignment results}
  \label{tab:joint_clip_results}
  \centering
  \begin{tabular}{@{}lcc@{}}
    \toprule
     & Top$1$-Acc.$\uparrow$ & Top$3$-Acc.$\uparrow$ \\ %
    \midrule
    Ours (TRELLIS SLats) & $81.31$ & $91.06$ \\ %
    Ours (both SLats) & $81.88$ & $91.41$ \\ %
    Ours (AR default order) & $84.68$ & $91.10$ \\ %
    Ours (AR BFS) & $\mathbf{85.90}$ & $\mathbf{92.06}$ \\ %
    Ours (AR DFS) & $84.61$ & $90.93$ \\ %
    Ours (AR z-y-x) & $74.24$ & $86.29$ \\ %
  \bottomrule
  \end{tabular}
\end{table}

\section{Rig Label Dataset Preparation Details}\label{sec:label_data_details}

In this section, we provide the full prompt details regarding the LLM based filtering and standardization of the labels provided with the Anymate dataset \cite{deng2025anymate}.
The filtering resulted in approximately $153$k samples with informative labels in the training split and $\sim  3.5k$ in the test-split.

The full prompt for processing the raw labels is provided in Listing \ref{lst:prompt}, and examples of labels before and after the processing are shown in Table~\ref{tab:labels_before_after}.

\section{Broader Societal Impact}
\label{sec:impact}
This work focuses on object-level 3D animation and does not directly involve sensitive personal data or decision-making systems. It may reduce manual effort in animation workflows, with potential misuse limited to generating misleading animated 3D content.

\newpage

\begin{table}[h]
  \caption{Examples of joint label preprocessing.}
  \label{tab:labels_before_after}
  \centering
  \begin{tabular}{@{}ll@{}}
    \toprule
     Unprocessed label & Processed label \\
    \midrule
    \texttt{metarig--spine.1} & \texttt{Spine1} \\
    \texttt{Armature--mixamorig:LeftForeArm} & \texttt{LeftForeArm} \\
    \texttt{Armature--Mutant:LeftToeBase} & \texttt{LeftToeBase} \\
    \texttt{GLTF\_created\_0--shoulder.R\_Skeleton\_14} & \texttt{RightShoulder} \\
    \texttt{Bip001--Bip001\_L\_UpperArm} & \texttt{LeftUpperArm} \\
    \texttt{group--pasted\_\_Neck} & \texttt{Neck} \\
    \texttt{SK\_Female\_Rig\_V1\_MainC--SK\_Female\_Rig\_V1\_SH\_lShoulderJ} & \texttt{LeftShoulder} \\
  \bottomrule
  \end{tabular}
\end{table}

\begin{lstlisting}[caption={Joint label filtering prompt},label={lst:prompt}]
"### ROLE\n"
"You are an intelligent AI assistant for computer graphics, specialized in rigging."
"Your goal is to standardize joint labels for professional 3D pipelines (Maya/Blender/Unity)."
"Follow the user's requirements carefully and make sure you understand them. \n\n"

"### TASK\n"
"Process a list of joint labels separated by ' || '. For each label:\n"
"1. Remove technical prefixes (e.g., 'mixamorig:', 'Armature:', 'Bind_', 'jnt_', 'bn_').\n"
"2. Standardize to PascalCase (e.g., 'lower_arm_L' -> 'LowerArmL').\n"
"3. Translate non-English anatomical terms to English (e.g., 'Cuisse' -> 'Thigh', 'Brazo' -> 'Arm'). **Do not guess; if unsure, keep the original.**\n"
"4. **PRESERVE** numbers and side indicators (e.g., 'Thigh1', 'Arm.R', 'LegC'). Do not strip these.\n\n"

"### CRITICAL LOGIC: INFORMATIVE VS UNINFORMATIVE\n"
"- **Informative:** Labels containing recognizable body part or object part names (Thigh, Finger, Spine, Wheel, Trigger, etc.).\n"
"- **Uninformative:** Labels that are generic (e.g., 'Joint1', 'Bone.001', 'Obj_42').\n"
"- :warning: **ACTION:** If **ANY** single label in the provided group is 'Uninformative', "
"you must return 'uninformative' as the entire result for that group.\n"
"- Use 'uninformative' only for strictly generic names. If a label contains a hint of anatomy (e.g., 'Arm_L_003'), treat it as informative.\n\n"

"### OUTPUT FORMAT\n"
"- Return ONLY a raw JSON list of strings.\n"
"- No conversational filler, no markdown code blocks, just the JSON.\n\n"
"- **IMPORTANT: Your response must start with [ and end with ]. Do not include any text before or after the JSON array.**"

"### EXAMPLE\n"
"Input: 'mixamorig:LeftBrazo || Armature:Cuisse_01 || Joint_02'\n"
"Output: [\"uninformative\"]\n"
"Input: 'mixamorig:LeftBrazo || Armature:Cuisse_01'\n"
"Output: [\"LeftArm\", \"Thigh01\"]"
\end{lstlisting}

\end{document}